\documentclass{article}
\usepackage[utf8]{inputenc}

\usepackage[english]{babel}
\usepackage{amsmath}
\usepackage{graphicx}
\usepackage[colorlinks=true, allcolors=blue]{hyperref}
\usepackage{amsthm}
\usepackage{comment}
\usepackage{enumitem}
\usepackage{mathtools}
\usepackage{amsfonts}
\usepackage{amssymb}
\usepackage {tikz}
\usetikzlibrary {positioning}
\tikzset{main node/.style={circle,fill=blue!40,draw,minimum size=1.5cm,inner sep=0pt},}
\tikzset{small dot/.style={fill=black,circle,scale=1.5}}
\tikzset{small dot two/.style={fill=red,circle,scale=1.5}}

\tikzset{small dot blue/.style={fill=blue!40,circle,scale=1.5}}
\usepackage{mathabx}

\tikzset{small dot 3/.style={fill=blue!40,circle,scale=1.25}}

\usetikzlibrary{arrows}

\DeclareMathOperator\supp{supp}

\usetikzlibrary{patterns}

\tikzset{try dot/.style={fill=black,circle,scale=1}}

\tikzset{try dot 2/.style={fill=blue!40,circle,scale=1}}

\usepackage{authblk}

\title{Majorization requires infinitely many second laws}
\author{Pedro Hack,  Daniel A. Braun, Sebastian Gottwald}
\date{ }

\newtheorem{proposition}{Proposition}

\newtheorem{corollary}{Corollary}

\newtheorem{defi}{Definition}

\newtheorem{theorem}{Theorem}

\usepackage{authblk}
\usepackage{blindtext}

\begin{document}

\maketitle

\begin{abstract}
Majorization is a fundamental model of uncertainty with several applications in areas ranging from thermodynamics to entanglement theory, and constitutes one of the pillars of the resource-theoretic approach to
physics. Here, we improve on its relation to measurement apparatuses. In particular, after discussing what the proper notion of second law in this scenario is, we show that, for a sufficiently large state space, any family of entropy-like functions constituting a second law must be countably infinite. Moreover, we provide an analogous result for a variation of majorization known as thermo-majorization which, in fact, does not require any constraint on the state space provided the equilibrium distribution is not uniform. Lastly, we discuss the applicability of our results to molecular diffusion and catalytic majorization. In this regard, we consider a variation of majorization used in plasma physics as a model of molecular diffusion and show that no finite family of entropy-like functions constituting a second law of molecular diffusion exists. Moreover, we
show how our results are useful when dealing with a conjecture regarding catalytic majorization (i.e. trumping). In particular, we show that the sort of characterizations of trumping that have been considered before require an infinite family of real-valued functions.
\end{abstract}


\section{Introduction}

Entropy was conceived as a fundamental tool in the study of transitions in thermodynamic systems. The fundamental notion that entropy aims to capture is that of disorder, or probability concentration. Although there is a long tradition of equating entropy and disorder, recently, another notion of disorder has gained momentum in several areas of physics, namely majorization \cite{sagawa2022entropy,nielsen2002introduction,nielsen2002quantum,gour2018quantum,lostaglio2019introductory}, a preorder defined on finite probability distributions. Despite the fact that majorization was introduced as a tool to analyze economic inequality \cite{dalton1920measurement,marshall1979inequalities}, it has a long tradition in physics which goes back to Ruch and Mead \cite{ruch1975diagram,ruch1976principle,mead1977mixing}, Alberti and Uhlmann \cite{alberti1981dissipative,alberti1982stochasticity} and Zylka \cite{zylka1985note}.\footnote{Note that one of the original motivations that led Ruch \cite{ruch1975diagram} to the study of majorization were the partitions of natural numbers. The study of these has been largely overlooked, with only a few more recent papers \cite{kirwan2016doppelganger,seitz2022mixed} being concerned with issues like the degeneracy in the Shannon entropy of partitions.}
One of the main physical motivations for majorization is the study of systems where the thermodynamical limit does not hold (see, for example, \cite{horodecki2013fundamental}). Moreover, majorization has found applications in the study of several physical phenomena, like the transitions in thermodynamic systems that are out of equilibrium and isolated \cite{ruch1976principle,horodecki2013fundamental,brandao2015second} or systems that obey a master equation. This has resulted, for instance, in the statement of the \emph{principle of increasing mixing character} \cite[Part B]{ruch1975diagram}. Furthermore, it has been proven to capture the conversion between entangled pure quantum states by local operations and classical communications \cite{nielsen1999conditions,jonathan1999entanglement,nielsen2002introduction}. More generally, majorization is also one of the key elements underlying the modern approach to several physical phenomena which is based on order structures and is known as \emph{resource theory} \cite{gour2015resource,lostaglio2019introductory,winter2016operational}.

The main aim of this work is to provide insight regarding the relation between entropy and majorization. In particular, we study how the order relationship for state transitions implied by the entropy concept as expressed by the second law can be carried over to the setting of majorization. First of all, entropy loses its privileged position and becomes just one element in a family of measurements that characterize the majorization preorder. Crucially, the choice of such a family is not unique.
Second, the property of being a second law could, in principle, be assigned to the entire family as a whole or to each member of the family individually. In fact, we can think of the second law as simply a numerical characterization of the transitions allowed by majorization or as a family of competing optimization principles where transitions are allowed only if all of them agree. 
In this paper, we consider the strongest of these requirements in that we study the case where each member of a family represents a second law that distinguishes reversible and irreversible transitions and, all together, they distinguish impossible transitions as well. (It will become clear later on why we cannot ask for all functions to individually distinguish all impossible transitions.) That is, we study families of optimization principles that characterize majorization. Our main result consists in showing that under these assumptions one needs infinitely many entropy-like functions in order to obtain the analogue to the second law when the underlying disorder model is assumed to be given by majorization.

Moreover, we show that an analogous conclusion holds for $d$-majorization \cite{joe1990majorization,gour2015resource,lostaglio2019introductory}, also known as \emph{thermo-majorization} in resource theory, which captures disorder with respect to arbitrary distributions $d$ in contrast to disorder with respect to the uniform distribution as in regular majorization. Lastly, we show how our results apply to both molecular diffusion and catalytic majorization.

In order to achieve our main goal, we start in Section \ref{gambling} by giving an intuitive introduction to majorization in terms of gambling. Right after, in Section \ref{example}, we highlight the difference between majorization and entropy via the familiar example of molecular diffusion. We follow this by discussing, in Section \ref{second law}, what we mean by a \emph{second law} in the context of majorization.
We use
Following
these considerations, in Sections \ref{main sect}  and \ref{main sect II}, we prove our main results regarding the second law(s) for majorization and $d$-majorization, respectively. We conclude in Section \ref{application} by discussing the applicability of our results to the study of both molecular diffusion and catalytic majorization.

\section{Gambling, majorization, and uncertainty}
\label{gambling}

Majorization constitutes a fundamental notion of disorder that can be interpreted in terms of gambling. In this section, we follow the intuitive approach in \cite{hack2022classification}. (A similar interpretation can be found in \cite{brandsen2022entropy}.) 

In order to expose majorization from an intuitive point of view, let us consider a casino owner that intends to incorporate a new game to the casino. All the games under consideration follow the same mechanism: bets are placed before the realization of a random variable and the gamblers win whenever they predicted the outcome properly. For simplicity, we assume the casino owner only considers games whose outcomes belong to some finite set $\Omega$. In order to choose a game, the casino owner ought to consider the possible edges a gambler can obtain. 
A gambler may bet on any proper subset of $\Omega$, that is, on any non-zero number $i<|\Omega|$ of outcomes. 
Assuming a game can be characterized by a probability vector $p$ on $\Omega$, where $p_n$ is the probability that option $n$ realizes, the highest probability of winning (the edge) a gambler can have when betting on $i$ outcomes is given by the sum of the $i$ largest components of $p$.
Hence, the casino owner prefers a game characterized by $p$ over another characterized by $q$ whenever $ \sum_{n=1}^{i} p^{\downarrow}_n \leq \sum_{n=1}^{i} q^{\downarrow}_n$ for all $i<|\Omega|$, where $p^{\downarrow}$ represents a rearrangement of $p$ with the components being ordered in non-increasing fashion. This defines the majorization preorder $\preceq$ on the set $\mathbb P_\Omega$ of probability distributions on $\Omega$. Formally,
\begin{equation}
\label{majo}
    p \preceq q \ \iff \ u_i(p) \leq u_i(q) \ \  \forall i\in \{1,..,|\Omega|-1\} \, ,
\end{equation}
where $u_i(p) \coloneqq \sum_{n=1}^{i} p^{\downarrow}_n$. Hence, if $\preceq$ is interpreted as ordering preferences, one could say majorization represents the preferences that a rival casino owner (who wishes the first one to enter bankruptcy) would recommend when inquired about what new game the first owner should incorporate. To illustrate the simplest example, we can consider games with two outcomes, which we can take w.l.o.g. to be coin tosses. In this scenario, all games are comparable, with balanced coins being the best from casino owner's perspective and the owner's preference decreasing as the coin's bias increases.


Majorization represents a fundamental notion of uncertainty, in the sense that $p\preceq q$ means that $p$ contains more uncertainty than $q$, because if summing the $i$ largest probabilities in $q$ is always larger than summing the $i$ largest probabilities in $p$ then $q$ must be more concentrated and $p$ less biased, i.e. `closer` to uniform. In fact, one of many equivalent characterizations of \eqref{majo}, based on moving pieces of probability, makes this interpretation explicit. In particular, $p\preceq q$, i.e. \emph{$q$ majorizes $p$}, if and only if $p$ can be obtained from $q$ by only moving pieces of probability from higher values $q_n$ to lower values $q_m$. Precursors of this idea can be found in the early 20th century economics literature about income inequality and wealth concentration \cite{dalton1920measurement,lorenz1905methods, pigou1912wealth}. From this characterization, it is clear that the uniform distribution on $\Omega$ is the smallest probability vector in $\mathbb P_\Omega$ with respect to $\preceq$, containing the most uncertainty and thus resulting in the lowest possible edge a gambler can have.

There is, however, a downside of using majorization \eqref{majo} as a way to order distributions with respect to their uncertainty: it only defines a \emph{preorder} on $\mathbb P_\Omega$. This means that, in contrast to \emph{total orders}, one cannot compare any two distributions, simply because $p\preceq q$ requires all the functions $u_i$ in \eqref{majo} to satisfy $u_i(p)\leq u_i(q)$, which is certainly not true for many, in fact uncountably many, distributions in general (c.f. examples in the following sections). Instead, the most common way to quantify uncertainty is by comparing the values of a single function, usually Shannon entropy $H$ given by $H(p)\coloneqq - \sum_i p_i \log p_i$, which conveniently allows to compare any two distributions on the same space. In the following section we briefly illustrate the differences between majorization and Shannon entropy by using the familiar example of molecular diffusion.

\section{Example: Molecular diffusion}
\label{example}

\begin{figure}[!tb]
\centering
\begin{tikzpicture}
\node[rounded corners, draw, text height = 2cm, minimum width = 9cm] {};
\node[rounded corners, draw, text height = 2cm, minimum width = 9cm,yshift=-3.5cm] {};

\draw [dashed] (-1.5,1) -- (-1.5,-1);
\draw [dashed] (1.5,1) -- (1.5,-1);
\draw [dashed] (-1.5,-2.5) -- (-1.5,-4.5);
\draw [dashed] (1.5,-2.5) -- (1.5,-4.5);

\node[small dot blue] at (-3,0) {};
\node[small dot blue] at (-4,0.75) {};
\node[small dot blue] at (-2.5,0.75) {};
\node[small dot blue] at (-3.5,-0.85) {};
\node[small dot blue] at (-1.8,-0.3) {};
\node[small dot blue] at (-3.8,-0.3) {};

\node[small dot blue] at (-3+3,0) {};
\node[small dot blue] at (-4+3,0.75) {};
\node[small dot blue] at (-2.5+3,0.75) {};
\node[small dot blue] at (-3.5+3,-0.85) {};
\node[small dot blue] at (-1.8+3,-0.3) {};
\node[small dot blue] at (-3.8+3,-0.3) {};

\node[small dot blue] at (-3,0-3.5) {};
\node[small dot blue] at (-4,0.75-3.5) {};
\node[small dot blue] at (-2.5,0.75-3.5) {};
\node[small dot blue] at (-3.5,-0.85-3.5) {};
\node[small dot blue] at (-1.8,-0.3-3.5) {};
\node[small dot blue] at (-3.8,-0.3-3.5) {};
\node[small dot blue] at (-2.3,-4) {};
\node[small dot blue] at (-2.3,-3.3) {};

\node[small dot blue] at (-3+3,0-3.5) {};
\node[small dot blue] at (-4+3,0.75-3.5) {};

\node[small dot blue] at
(-3.5+6,-3.3) {};
\node[small dot blue] at (-1.8+6,-0.3-3.5) {};

\draw [-implies,double equal sign distance] (-1,-1.2) -- (-1,-2.2);

\draw (-1.3,-2) -- (-0.7,-1.4);

\draw[thick,->]
    (1,-1.2) -- (1,-2.2) node[midway,sloped,yshift=-3mm,font=\fontsize{15}{15}\selectfont, thick, rotate=-90] {$H$};

\end{tikzpicture}
\caption{Simple system of two molecular states for a gas in a box where the increase in entropy $H$ allows transitions which are not possible. While the top state has lower entropy than the bottom one, we do not expect to observe transitions from the top one to the bottom one.}
\label{gas trans}
\end{figure}
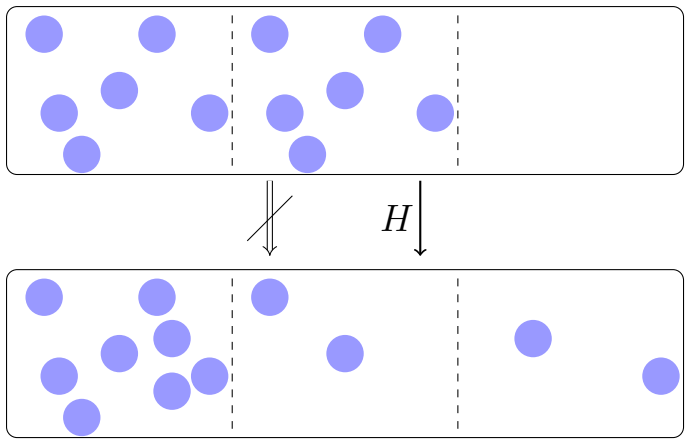

The distinction between entropy and majorization was nicely pointed out by Mead \cite{mead1977mixing} in a molecular diffusion setting. 
Consider a gas in a box that is divided into three compartments by two walls, and assume that we prepare it in such a way that half of the molecules are in the first compartment and the other half of them are in the second compartment. Another possible state of the system would be one where $2/3$ of the molecules are in the first compartment and $1/6$ of them occupy the second and the third compartments each (see Figure \ref{gas trans} for a representation of both states). In this scenario, the following is a natural question:
\emph{If we prepare the system in the first state and eliminate the walls that separate the compartments, will it spontaneously evolve and reach the second state?} 
From a molecular diffusion point of view, we would say that this is not the case, since the molecules would have a higher concentration in the first compartment than before. However, the Shannon entropy of the second state is higher than that of the first one. Hence, if we consider the increase of entropy as a necessary and sufficient criterion for a transition to take place, the transition is supposed to happen, contradicting molecular diffusion. Despite not settling it fully, majorization improves on this issue, as we will see in the following.


The fundamental problem with the example above is that the interchange of molecules between adjacent compartments tends towards equalizing the number of them in each compartment, resulting in an increase in uncertainty represented by the corresponding distribution. However, an increase in entropy may occur in a situation where the dissimilarity in the number of molecules between adjacent compartments is increased instead, as the simple example above shows.
Therefore, molecular diffusion can be better captured using majorization \cite{marshall1979inequalities,arnold2018majorization}. If probability distributions on $\Omega$ represent the fraction of molecules in each compartment (in our example, $|\Omega|=3$ and $p_i=n_i/N$ with $N$ the total number of molecules in the box and $n_i$ being the occupation number of the compartment $i$ for $i=1,2,3$), then we can use majorization as a model for molecular diffusion. In particular, we can identify the possibility of a thermodynamic transition from a distribution $q$ to a distribution $p$ with the majorization relation $p \preceq q$, where we call transitions \emph{thermodynamic} whenever they are allowed by molecular diffusion. For example, majorization allows to discard the transition between the states shown in Figure \ref{gas trans} (as they are not comparable with respect to $\preceq$), in accordance with molecular diffusion. In fact, this kind of considerations led Ruch \cite{ruch1975diagram,ruch1976principle} to the substitution of the principle of increasing entropy by the \emph{principle of increasing mixing character}, which is based on majorization. Despite this improvement, it is known that majorization does not perfectly capture molecular diffusion, an issue to which we will return in Section \ref{application} (where we propose a variation of majorization as a model of molecular diffusion and show how our results for majorization apply there).

\section{The second law in majorization-based thermodynamics}
\label{second law}

The use of order structures like majorization in thermodynamics has a long tradition. Such approaches can be traced back to Carathéodory \cite{caratheodory1909untersuchungen}, who initiated a long tradition \cite{cooper1967foundations,landsberg1970main,giles2016mathematical} that crystallized in the remarkable work by Lieb and Yngvason \cite{lieb1999physics}. (See also \cite{lieb2013entropy,lieb2014entropy}.) All these have a common methodological trait, namely, they begin by assuming some kind of order structure representing state transitions and intend to characterize it in terms of real-valued functions, or \textit{measurements}. 

In contrast to majorization, \cite{lieb1999physics} is focused on some subsets where the studied order structure is a total order and therefore allows a single measurement to characterize the transitions. For $|\Omega|=2$, we have  
\begin{equation}
\label{second law dim2}
    p \preceq q \ \iff \ H(p) \geq H(q).
\end{equation}
Hence, in this scenario, comparing entropy values is equivalent to majorization and we have a situation analogous to that in \cite{lieb1999physics}.
However, in the case of $|\Omega| \geq 3$, although every pair of distributions can be compared in terms of entropy, some of them are not comparable with respect to majorization. An example of this is the pair 
$p=(2/3,1/6,1/6, 0,\dots,0)$,
$q=(1/2,1/2,0,\dots,0)$,
in analogy to the example above, where each distribution has $|\Omega|-3$ zeros. In fact, in any $\mathbb P_\Omega$ with $|\Omega| \geq 3$, there are infinitely many distributions that are incomparable with respect to $\preceq$. 


Whereas a single function is insufficient to characterize majorization, \emph{a family} of functions can be used instead, also known as a \emph{multi-utility} in the literature on mathematical economics \cite{evren2011multi}. A multi-utility is a family $(f_i)_{i\in I}$ of real-valued functions such that each member of the family is monotonic with respect to the preorder, and, importantly, it allows to decide whether $p\preceq q$ by comparing the corresponding function values, i.e.,
\begin{equation} \label{multiutil}
p\preceq q \ \iff \ f_i(p)\leq f_i(q) \quad \forall i\in I.
\end{equation}
An example is the family \smash{$(u_i)_{i=1}^{|\Omega|-1}$} that we used to define majorization in \eqref{majo}. However, in order to characterize the possible majorization \emph{transitions} in the same spirit as the second law of thermodynamics, it turns out that the functions in such a family are required to be \emph{strictly} monotonic with respect to $\preceq$. As we will see below, this allows to faithfully represent irreversibilities in the majorization relation, as the single function used in \cite{lieb1999physics} does. Such a family can then be considered a generalization of entropy, as we discuss in more detail in the following.



\subsection{Two notions of second law}

In this section, we will recall two notions of the second law that have been historically considered and we will argue why we are interested in the stronger one, which we will consider in the remainder on this work. Importantly, we will note that these notions coincide whenever the state space is connected (i.e. given any pair of states there is always a process connecting them either from one to the other or the other way around).

The first notion of second law would be the one where we only consider entropy as a numerical characterization of the transitions a system may undertake. In this regard, historically, some common requirements connecting entropy and spontaneous processes are the following:
\begin{enumerate}[label=(\itshape\roman*)]
\item if a transition is reversible, then entropy remains unchanged, 
\item if a transition is irreversible, then entropy increases, and
\item if a transition is impossible, then entropy decreases.
\end{enumerate}
Note that $(i)$-$(iii)$ allow to conclude from the entropy values of two distributions, whether a certain transition would be allowed or not.

In this approach to the second law,
the requirements on a function to be called \emph{entropy} have varied throughout the development of physics. The first requirement on entropy was to increase whenever a process is possible. (This would correspond to $(i)$ together with a weak version of $(ii)$ where we exchange \emph{increase} by \emph{not decrease}.) Although such a weak requirement alone would also include trivial functions like constant ones, it also implies that a decrease in entropy directly rules out a process. A more demanding definition would consist precisely of properties $(i)$ and $(ii)$, with entropy also allowing us to distinguish reversible from irreversible transitions in this scenario. In fact, this constitutes the framework in the work by Lieb and Yngvason \cite{lieb1999physics}, where they are concerned with finding some function (which they call \emph{monotonic}) that fulfills properties  $(i)$ and $(ii)$. (They refer to the fulfillment of such properties as the \emph{entropy principle}.) Lastly, by only considering certain state spaces, they end up being able to fully characterize the transitions via a single function (analogous to the characterization of majorization via Shannon entropy in \eqref{second law dim2}) and, hence, also fulfilling $(iii)$ (since, in this setup, the only impossible transitions are the inverse of reversible ones and, thus, $(ii)$ implies $(iii)$).

Although Lieb and Yngvason do not deal with incomparabilities, as we will later see, it is the study of transition systems containing pairs of elements that are \emph{not} related by any process what is relevant for our discussion here, since, there, the two notions of second law we consider here diverge. Returning to the first notion of second law, and in the spirit of simply considering the second law as a numerical characterization of the allowed transitions, multi-utilities like the one in \eqref{majo} are a natural generalization of $(i)-(iii)$ when incomparabilities are present. In fact, such families of functions are commonplace in the resource-theoretic literature on the second law in majorization-based thermodynamics \cite{brandao2015second,goold2016role,lostaglio2019introductory,lostaglio2022continuous}.

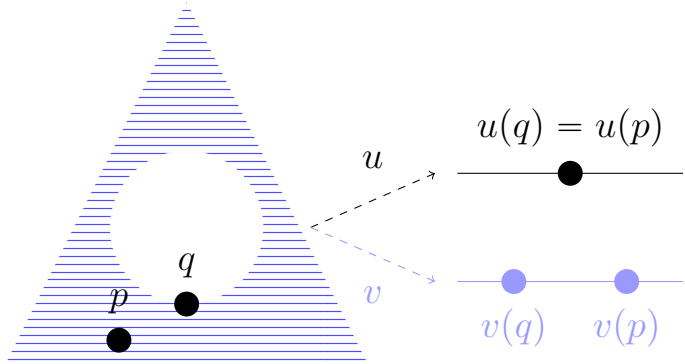
\begin{figure}[!tb]
\centering
\begin{tikzpicture}[scale=0.6]
\fill [white, pattern=horizontal lines, pattern color=blue!70] (-2,4)--(-4-2,-4)--(4-2,-4)--cycle;
\path [fill=white] (-2,-1) circle (1.7cm);

\draw [->,dashed] (0.75,-1) -- node[above =0.3cm, black,font=\fontsize{15}{15}\selectfont] {\textbf{$u$}} (3.5,0.2);
\draw [->,color= blue!40,dashed] (0.75,-1) -- node [below =0.3cm, color= blue!40,font=\fontsize{15}{15}\selectfont] {\textbf{$v$}} (3.5,-2.2);

\draw [-] (4,0.2) -- (9,0.2);
\draw [-,color= blue!40] (4,-2.2) -- (9,-2.2);

\node[try dot,label={[anchor=north,above=0.5mm, thick, font=\fontsize{15}{15}\selectfont, thick]90:$q$}] at (-2,-2.7) (1) {};

\node[try dot,label={[anchor=north,above=0.5mm, thick, font=\fontsize{15}{15}\selectfont, thick]90:$p$}] at (-3.5,-3.5) (2) {};

\node[try dot,label={[anchor=north,above=0.5mm, thick, font=\fontsize{15}{15}\selectfont, thick]90:$u(q)=u(p)$}] at (6.5,0.2) (3) {};

\node[try dot 2,label={[anchor=south,below=0.5mm, thick, font=\fontsize{15}{15}\selectfont, thick, color= blue!40]270:$v(q)$}] at (5.25,-2.2) (4) {};

\node[try dot 2,label={[anchor=south,below=0.5mm, thick, font=\fontsize{15}{15}\selectfont, thick, color= blue!40]270:$v(p)$}] at (7.75,-2.2) (5) {};

\end{tikzpicture}
\caption{Schematic representation of the difference between a function $u$ that belongs to a multi-utility and one $v$ that fulfills property $(ii)$ in the $2$-simplex. In particular, given $p,q \in \mathbb P_\Omega$ such that $q \prec p$ and $q$ is minimal in some subset $B \subseteq \mathbb P_\Omega$ (striped area), $v$ is only minimized by minimal elements of $\preceq$ like $q$, while $u$ may be minimized by non-minimal elements like $p$.}
\label{fig:property ii}
\end{figure}

In the spirit of \cite[Chapter I.8]{landau2013statistical} (Jaynes' maximum entropy principle \cite{jaynes1957information} follows an analogous argument), and as an alternative notion, we think of the second law as an optimization principle. By this, we mean that entropy is a quantity that always increases and, hence, whose maximization over some constrained region (for instance, a region where energy is fixed to a certain value) yields an equilibrium state. As we already stated, in case there are no incomparabilities, it is easy to see that both notions of second law are equivalent. However, the individual functions in the natural generalization of the first notion of second law, i.e. the functions in some multi-utility, cannot be interpreted as optimization principles if they do not individually fulfill $(ii)$. This is the case since, for each function in a multi-utility $u: \mathbb P_\Omega \to \mathbb R$ not fulfilling $(ii)$, we can find some region such that, when optimized over it, we obtain non-equilibrium states. (If $u$ does not fulfill $(ii)$ for some pair $p,q \in \mathbb P_\Omega$ with $p \prec q$, then, since $\mathrm{argmax}_B f = B$, a trivial instance of this is the set $B = \{p,q\}$, where we maximize $f = -u$ since the equilibrium states in $B$ are the lowest in $\preceq$  by \eqref{majo}. See Figure \ref{fig:property ii} for a visual representation using a non-trivial subset $B \subseteq \mathbb P_\Omega$ with $|\Omega|=3$.) However, if we require the functions involved in a multi-utility to individually fulfill $(ii)$, then we can interpret majorization as a family of competing optimization principles where the only allowed transitions are those in which they all agree. (It should be added that, given the \emph{entropy principle} in \cite{lieb1999physics}, we can also think of $(ii)$ as a necessary condition for a function to be called an \emph{entropy}.) This generalization of the second law is the one which we will be considering in what follows. We will refer to it as a \emph{family of second laws}.

\subsection{The families of second laws}

Formally, for any state space $\mathcal S$, we call a tuple $\mathcal T = (T_1,T_2,T_3)$ a set of \emph{possible transitions} on $\mathcal S$, if $\mathcal T$ forms a disjoint partition of $\mathcal S^2 = \mathcal S\times\mathcal S$. More precisely, $T_1$ and $T_2$ are interpreted as sets of state transitions $(s,t)$ from $t$ to $s$ corresponding to \emph{reversible} and \emph{irreversible transitions}, respectively, and $T_3 = (T_1\cup T_2)^c$ are thus the \emph{impossible transitions} $(s,t)$ from states $t$ to $s$. In our case, the states are simply given by probability distributions $p$ and $q$, i.e. $\mathcal S$ is the probability simplex over $\Omega$. 

A simple way to generate such a $\mathcal T$ is, for example, given by a preorder $\preceq$ on $\mathcal S$. Namely, we can define the set of possible transitions $\mathcal T_\preceq = (T_1,T_2,T_3)$ \emph{induced by $\preceq$}, by
\begin{align*}
T_1 & = \{(s,t) \in \mathcal S^2\, | \, s \preceq t  \text{ and } t\preceq s \text{, denoted by }s\sim t\},\\
T_2 & = \{(s,t) \in \mathcal S^2\, | \, s \preceq t  \text{ and } t\not \preceq s \text{ (denoted by }s\prec t\text{)}\},\\
T_3 & = \{(s,t) \in \mathcal S^2\, | \, s \not\preceq t\} = T_2^\top {\cup}\, U,
\end{align*}
where $T_2^\top = \{(s,t) \, | \, t\prec s\}$, and $U$ denotes the set of $(s,t)$ where $s$ and $t$ are uncomparable (also called \emph{incomparable}), i.e.~where neither $s\preceq t$ nor $t\preceq s$, sometimes denoted by $s\bowtie t$. Note that the disjoint union $T_3 = T_2^\top {\cup}\, U$ represents the two scenarios under which a transition from $t$ to $s$ is impossible: either there is an irreversible transition from $s$ to $t$, i.e.~$(t,s)\in T_2$ which means $(s,t)\in T_2^\top$, or $s$ and $t$ are incomparable, i.e.~$(s,t)\in U$.

Analogous to the properties $(i)$-$(iii)$ of entropy above, we now say that a function $f$ on $\mathcal S$ is a \emph{second law} for a set of possible transitions $\mathcal T$ if $f$ satisfies the following requirements:
\begin{enumerate}[label=(\itshape\roman*)]
\item $f(s) = f(t)$ for all $(s,t)\in T_1$,
\item $f(s) > f(t)$ for all $(s,t)\in T_2$, and
\item $f(s) < f(t)$ for all $(s,t)\in T_3$.
\end{enumerate}
By definition, entropy is then a second law for spontaneous processes in statistical mechanics. However, in general, an arbitrary set of possible transitions is not guaranteed to have a second law, simply because a single function on $\mathcal S$ is not enough to specify an arbitrary partition on $\mathcal S^2$ using $(i)$-$(iii)$. In fact, this problem already exists in the much smaller class of possible transitions induced by a preorder, as follows directly from our main results about majorization in the following sections. 

Notice the asymmetry in the definition of $\mathcal T_\preceq$ for a general preorder $\preceq$, specifically in the sets $T_2$ and $T_3$. If $\preceq$ is a total order, then $U$ is empty and thus $T_3=T_2^\top$, which corresponds to the symmetry of properties $(ii)$ and $(iii)$ of a second law. However, if $\preceq$ is a non-total preorder, then the existence of a single second law $f$ would entail that $f(t)>f(s)$ whenever $(s,t)\in T_3$, in particular, this is the case when the transition from $t$ to $s$ is impossible because $t$ and $s$ are incomparable, i.e.~when $(s,t)\in U$, which is then indistinguishable from the case when $(t,s)\in T_2$, i.e. when there is an irreversible transition from $s$ to $t$. 

In order to resolve this issue, we are introducing the concept of a \emph{family of second laws} $(f_i)_{i \in I}$, which is a straightforward generalization of the above notion of a single second law, specialized to the class of transitions induced by a preorder $\preceq$. In particular, in contrast to a single second law, it allows to distinguish the sets $U$ and $T_2^\top$ by only comparing function values. First, we assume each member of such a family to be monotonically decreasing with respect to $\preceq$, i.e.~if $s\preceq t$ then $f_i(t)\leq f_i(s)$ for all $i$. Furthermore, the requirements $(i)$ and $(ii)$ for a single second law directly extend to a family of second laws by simply requiring them to hold for all members of the family. In the case of $(s,t)\in T_3$, however, we can require such a family now to be able to distinguish the cases when $(s,t)\in T_2^\top$ and $(s,t)\in U$. In particular, if $(s,t)\in T_2^\top$ then there is an irreversible transition from $s$ to $t$, i.e.~$t\prec s$, which is the property $(ii)$ but for the reversed order $(t,s)$. In the case of $(s,t)\in U$, i.e.~$s\not \preceq t$ and $t\not \preceq s$, we borrow the property from multi-utilities that there are indices $i$ and $j$ such that $f_i(s)<f_i(t)$ and $f_j(s)>f_j(t)$.

In summary, given a preordered space $(\mathcal S,\preceq)$, equipped with the induced set of possible transitions $\mathcal T_\preceq$, we call a family $(f_i)_{i \in I}$ of real-valued functions $f_i:\mathcal S\to \mathbb R$ a \emph{family of second laws for $\preceq$} if 
\begin{enumerate}[label=(\itshape\roman*)]
\item $\forall (s,t) \in T_1$ ($s\sim t$):  $f_i(s)=f_i(t)$ for all $i\in I$, 
\item $\forall (s,t) \in T_2$ ($s\prec t$): $f_i(s)>f_i(t)$ for all $i\in I$,
\item $\forall (s,t) \in U$ ($s\bowtie t$): $\exists i,j\in I$ with $f_i(s)<f_i(t)$ and $f_j(s)>f_j(t)$.
\end{enumerate}
Note that the negative of a family of functions with these properties, $(g_i)_{i \in I}$ with $g_i=-f_i$ for all $i \in I$, is known as a \emph{Richter-Peleg} or \emph{strict monotone} multi-utility associated to a preorder $\preceq$ \cite{alcantud2016richter,hack2022representing}. Hence, the existence of a family of second laws is equivalent to that of a strict monotone multi-utility.

\subsection{The families of second laws in the literature}

Although multi-utilities have been extensively used, they are not the only functional characterizations of transition systems in resource theory. For instance, it is common in the study of catalytic majorization (i.e. trumping, see \eqref{def trump}) to require functions to individually fulfill $(ii)$ \cite{turgut2007catalytic,klimesh2007inequalities,klimesh2004entropy}. (One can check \cite[p. 3]{muller2016generalization} for a short overview concerning the kind of functional characterizations used in resource theory and the definition of a \emph{complete} set of monotones.) We will return to this point in more detail in Section \ref{sec trumping}, where we show how our results on the families of second laws apply to a conjecture on trumping. Lastly, to avoid confusion regarding the study of thermo-majorization in resource theory, it should be emphasized that we do not include any functional restrictions on our family of second laws, we simply require them to be related with the order structure via the properties $(i)-(iii)$. This contrasts with other definitions of the second law in the literature \cite{lostaglio2019introductory,brandao2015second}, which tie the second laws closely to $\alpha$-Rényi divergences and show that $(i)-(iii)$ cannot be fulfilled without the presence of a catalyst. (See \cite[Section 2.3.2]{lostaglio2019introductory} for more details.)

In the following section, restricting ourselves to the case when the preorder is given by majorization, we show our main result that establishes a lower bound on the number of functions constituting such a \textit{family of second laws for disorder}, as we refer to $(f_i)_{i\in I}$ with the properties $(i)$-$(iii)$ in this case.

\section{An infinite family of second laws is needed to characterize majorization}
\label{main sect}

The lower bound on the number of functions needed to conform a family of second laws of disorder varies with the size of $\Omega$. In particular, in the case $|\Omega|=2$, then the lower bound is one, as stated in \eqref{second law dim2}, with negative entropy being an example of such a family. However, if $|\Omega| \geq 3$, then the functions in \eqref{majo} do not constitute a family of second laws of disorder, since they do not respect $(ii)$.\footnote{In fact, if $|\Omega| \geq 3$, it is easy to find, for all $i$ with $1 \leq i \leq |\Omega|-1$, an irreversible transition $p^i \prec q^i$ such that $u_i(p^i) = u_i(q^i)$. For instance, fixing some $p \in \mathbb P_\Omega$ with $p_i >p_{i+1}>0$ for all $i < |\Omega|$, we can define $p^i \coloneqq p$ for all $i$ with $1 \leq i \leq |\Omega|-1$ and $q^i$ with the components equal to $p$ except, if $i < |\Omega|-1$, for $(q^i)_{i+1}=p_{i+1}+ \varepsilon_i$ and $(q^i)_\Omega=p_\Omega- \varepsilon_i$, where $0<\varepsilon_i < \text{min} \{p_i-p_{i+1},p_\Omega \}$, and, if $i = |\Omega|-1$, for $(q^i)_{1}=p_{1}+ \varepsilon_{|\Omega|-1}$ and $(q^i)_{2}=p_{2}- \varepsilon_{|\Omega|-1}$, where  $0<\varepsilon_{|\Omega|-1} < \text{min} \{1-p_1,p_1-p_2\}$.} Therefore, the obvious question we have to answer is whether such a family even exists, and if so, what is the minimal number of members of such a family? In Theorem \ref{dim majo}, we answer both of these questions. More specifically, we show that, although countably infinite families of second laws of disorder do exist, there is no finite family of second laws whenever $|\Omega| \geq 3$.

\begin{theorem}
\label{dim majo}
If $|\Omega| \geq 3$, then the smallest family of second laws of disorder is countably infinite. 
\end{theorem}

\begin{proof}
The first thing we ought to notice is that a countably infinite family of second laws of disorder does indeed exist if $|\Omega| \geq 3$. In particular, as noted in \cite[Theorem 3.1]{alcantud2016richter} in a more general setting and for strict monotone multi-utilities, $(f_{i,n})_{1 \leq i \leq |\Omega|-1, n \geq 0}$ is a family of second laws of disorder, where $f_{i,n} \coloneqq H - q_n u_i$ and $(q_n)_{n \geq 0}$ is a numeration of the strictly positive rational numbers.

To conclude, we only need to show that no finite family of second laws of disorder exists if $|\Omega| \geq 3$. In order to do so, we will show there exists a subset $S \subseteq \mathbb P_\Omega$ such that $(S,\preceq_S)$ is order isomorphic to an ordered set $(X,\preceq')$ without finite families of second laws. That is, we will show that there exists a bijective map $f:S \to X$ such that, for all $p,q \in S$, $p \preceq_S q \iff f(p) \preceq' f(q)$, where $\preceq_S$ is the majorization relation restricted to $S$. In particular, we take $(X,\preceq')$ the ordered set defined in \cite[Theorem 2]{hack2022geometrical}, that is, $X \coloneqq \mathbb{R}\setminus\{0\}$ and $\preceq'$ is defined by
\begin{equation}
\label{simple partial}
    x \preceq' y \iff
    \begin{cases}
    |x| \leq |y| \text{, and } \\ sgn(x) \leq sgn(y)
    \end{cases}
\end{equation}
for all $x,y \in X$, where $|x|$ and $sgn(x)$ denote, respectively, the absolute value of $x$ and the sign of $x$, that is, $sgn(x) \coloneqq 1$ if $x> 0$ and $sgn(x) \coloneqq -1$ if $x<0$. (A representation of $(X,\preceq')$ can be found in Figure \ref{fig:counterex}.) Since the existence of a finite family of second laws of disorder for $|\Omega| \geq 3$ would imply that of a finite family of second laws for $(S,\preceq_S)$ and, hence, one for $(X,\preceq')$ (which does not exist, as shown in the proof of \cite[Theorem 3 $(ii)$]{hack2022geometrical} in the context of strict monotone multi-utilities), we obtain that no finite family of second laws of disorder exists.

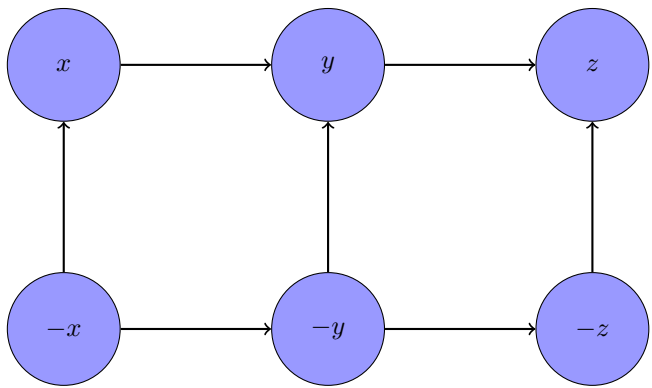
\begin{figure}[!tb]
\centering
\begin{tikzpicture}
    \node[main node] (1) {$x$};
    \node[main node] (2) [right = 2cm  of 1]  {$y$};
    \node[main node] (3) [right = 2cm  of 2]  {$z$};
    \node[main node] (4) [below = 2cm  of 1] {$-x$};
    \node[main node] (5) [right = 2cm  of 4] {$-y$};
    \node[main node] (6) [right = 2cm  of 5] {$-z$};

    \path[draw,thick,->]
    (4) edge node {} (5)
    (5) edge node {} (6)
    (1) edge node {} (2)
    (2) edge node {} (3)
    (4) edge node {} (1)
    (6) edge node {} (3)
    (5) edge node {} (2)
    ;
\end{tikzpicture}
\caption{Representation of the simple ordered set defined in \eqref{simple partial} that we use in Theorem \ref{dim majo} to show that no finite family of second laws of disorder exists if $|\Omega| \geq 3$. Note that, directed graphs can be used to represent orderings $\preceq$, such that $x\preceq y$ whenever there exists a path following the arrows from $x$ to $y$. We include three positive real numbers $0<x<y<z$ in the top line and their negatives $-x,-y,-z$ in the bottom line.}
\label{fig:counterex}
\end{figure}

Lastly, we construct $S \subseteq \mathbb P_\Omega$ and show it is order isomorphic to $(X,\preceq')$. We fix, for the moment, $|\Omega|=3$ and return to the general case later. Fix $\varepsilon, \gamma >0$ such that $\gamma < \varepsilon < \frac{1}{4}$ and take
\begin{equation} \label{def S}
\begin{split}
S &\coloneqq \{p_x, q_x\}_{x \in (\frac{1}{2}, \frac{1}{2} + \varepsilon - \gamma)}, \text{ where} \\
 p_x &\coloneqq (x, \frac{1}{4}+ \varepsilon-(x- \frac{1}{2}), \frac{1}{4}-\varepsilon) \text{ and} \\
 q_x &\coloneqq (x, \frac{1}{4}+ \varepsilon-(x- \frac{1}{2})- \gamma, \frac{1}{4}-\varepsilon + \gamma)
\end{split}
\end{equation}
for all $x \in (\frac{1}{2}, \frac{1}{2} + \varepsilon - \gamma)$.
Note $(p_x)_2 < (p_x)_1$ since $\frac{1}{4}+ \varepsilon-(x- \frac{1}{2}) < \frac{1}{2}-(x- \frac{1}{2})=1-x <x$ and $(p_x)_3 < (p_x)_2$ since $x < \frac{1}{2} + \varepsilon - \gamma < \frac{1}{2} + 2\varepsilon$. Hence, $p_x$ is non-increasingly ordered $p_x=p^{\downarrow}_x$ for all $x \in (\frac{1}{2}, \frac{1}{2} + \varepsilon - \gamma)$. Note, also, $(q_x)_2 < (q_x)_1$ since
$(q_x)_2=(p_x)_2- \gamma < (p_x)_2 < (p_x)_1=(q_x)_1$ and $(q_x)_3 < (q_x)_2$ since $x < \frac{1}{2} + (\varepsilon - \gamma) < \frac{1}{2} + 2(\varepsilon - \gamma)$. Thus, $q_x$ is non-increasingly ordered $q_x=q^{\downarrow}_x$ for all $x \in (\frac{1}{2}, \frac{1}{2} + \varepsilon - \gamma)$.
To obtain that $(S,\preceq_S)$ is order isomorphic to $(X,\preceq')$, it suffices to show that, for any pair $x,y \in (\frac{1}{2}, \frac{1}{2} + \varepsilon - \gamma)$ with $x <y$, we have $p_x \prec p_y$, $q_x \prec q_y$, $q_x \prec p_x$ and $p_x \bowtie q_y$. (See Figure \ref{fig:majo} for a representation of $(S,\preceq_S)$.) These relations are easy to see given that we have  $u_1(p_x)=u_1(q_x)=x < y =u_1(p_y) = u_1(q_y)$ and $u_2(q_x)=u_2(q_y)= \frac{3}{4}+ \varepsilon - \gamma < \frac{3}{4}+ \varepsilon =u_2(p_x)=u_2(p_y)$. As a result, given some bijection $g: (\frac{1}{2},\frac{1}{2} + \varepsilon - \gamma) \to (0,\infty)$, we can take 
 \begin{alignat*}{3}
    f:\text{ } &S &&\to &&\text{ }X\\
    &p_x &&\mapsto &&\text{ }g(x)\\
    &q_x &&\mapsto &&\text{ }-g(x)
\end{alignat*}
as order isomorphism.\footnote{Note that we could have constructed the order isomorphism from $S$ to some subset of $X$ (like the more natural one where $\frac{1}{2}<|x|<\frac{1}{2}+\epsilon-\gamma$) and the proof would work exactly the same (cf. \cite[Theorem 3 $(ii)$]{hack2022geometrical}). For simplicity, we follow \cite{hack2022geometrical} and use $X$. Similar considerations apply to the proofs of Theorems \ref{d-majo dim} and \ref{thm: small state space}.}

To finish, notice we can show the case $|\Omega|>3$ analogously to the case $|\Omega|=3$. We can simply add $|\Omega|-3$ zeros as last components for both $q_x$ and $p_x$ for all $x \in (\frac{1}{2}, \frac{1}{2} + \varepsilon - \gamma)$.
\end{proof}

In the following section, we show that an analogous conclusion to that of Theorem \ref{dim majo} holds for a variation of majorization known as $d$-majorization.

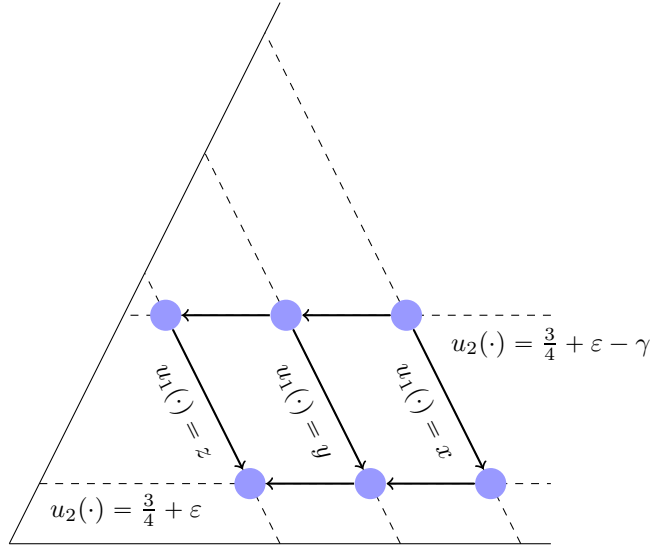
\begin{figure}[!tb]
\centering
\begin{tikzpicture}[scale=0.8]

\draw (-5,-5)--(4,-5);

\draw (-5,-5)--(-0.5,4);


\draw [dashed] (3.5,-5) node {} -- (-3/4,3/2+2);

\draw [dashed] (1.5,-5) node {} -- (-7/4,7/2-2);

\draw [dashed] (-0.5,-5) node {} -- (-11/4,5-11/2);

\draw [dashed] (4,-1.2) node[below =0.05 cm] {$u_2(\cdot)= \frac{3}{4}+ \varepsilon - \gamma$} -- (-3.1,-1.2);

\draw [dashed] (-9/2,-4) node[below right =0.03 cm] {$u_2(\cdot)= \frac{3}{4}+ \varepsilon $}-- (4,-4);


\node[small dot 3] at (-12/5,-1.2) (1u) {};

\node[small dot 3] at (-2/5,-1.2) (2u) {};

\node[small dot 3] at (8/5,-1.2) (3u) {};


\node[small dot 3] at (-1,-4) (1d) {};


\node[small dot 3] at (1,-4) (2d) {};

\node[small dot 3] at (3,-4) (3d) {};

\path[draw,thick,->]
    (1u) edge node [midway,sloped,below =0.05 cm] {$u_1(\cdot)=z$} (1d)
    (2u) edge node [midway,sloped,below =0.05 cm] {$u_1(\cdot)=y$} (2d)
    (3u) edge node [midway,sloped,below =0.05 cm] {$u_1(\cdot)=x$} (3d)
    (3d) edge node {} (2d)
    (2d) edge node {} (1d)
    (3u) edge node {} (2u)
    (2u) edge node {} (1u)
    ;

\end{tikzpicture}
\caption{Representation of $(S,\preceq_S)$ inside the $2$-simplex. We introduced $(S,\preceq_S)$ in Theorem \ref{dim majo} to show that no finite family of second laws of disorder exists if $|\Omega| \geq 3$. Again, we represent $\preceq_S$ as a directed graph, where $p\preceq_S q$ whenever there exists a path following the arrows from $p$ to $q$, and $\frac{1}{2}<x<y<z<\frac{1}{2}+\varepsilon-\gamma$.
From this representation, we can see immediately that $(S,\preceq_S)$ is order isomorphic to the partial order in Figure \ref{fig:counterex}, which is key in Theorem \ref{dim majo}. Note this figure also represents $(S_d,\preceq_{S_d})$, which we introduced in Theorem \ref{d-majo dim}, provided we think of it as the projection of $\mathbb P_\Omega$ where, for any distribution $q$, we have $q_m=d_m$ for $m=3,\dots,|\Omega|-1$.}
\label{fig:majo}
\end{figure}

\section{An infinite family of second laws is needed to characterize d-majorization}
\label{main sect II}

Similarly to how we interpreted majorization in Section \ref{gambling}, we can consider a variation of it which is known in the resource-theoretic literature as \emph{thermo-majorization} \cite{horodecki2013fundamental,gour2015resource,lostaglio2019introductory} and was originally called $d$-majorization in \cite{joe1990majorization}. The basic idea is that, instead of modelling general uncertainty as majorization does, we model uncertainty relative to a specific distribution. In several physical applications, the reference distribution is a Gibbs distribution \cite{horodecki2013fundamental,gour2015resource,lostaglio2019introductory}. This can be read directly in terms of our analogy with molecular diffusion, where majorization would model an isolated system, and $d$-majorization, with $d \in \mathbb P_\Omega$ being a Gibbs distribution, a system in contact with a heat bath. In the latter scenario, there is an energy function which influences molecules, making them not only diffuse between compartments but actually prefer some compartments over others. The presence of a non-constant energy function breaks the symmetry between compartments and results in a non-uniform equilibrium distribution. In particular, $d$ is usually taken to have the following form
\begin{equation}
\label{gibbs distri}
    d = \frac{1}{Z} \left( e^{-\beta E_1}, \dots, e^{-\beta E_{|\Omega|}}\right),
\end{equation}
where $E_i$ is the energy associated to compartment $i$, $\beta$ is the inverse temperature and $Z \coloneqq \sum_{i=1}^{|\Omega|}  e^{-\beta E_i}$ is the partition function.
Such an approach was also pioneered by Ruch and constitutes what is called the \emph{principle of decreasing mixing distance} \cite{ruch1976principle,ruch1978mixing}.
This principle states that, during the evolution of the system, its distribution becomes closer to the Gibbs distribution, with the notion of \emph{close} being given by $d$-majorization.

Before stating the main result of this section, let us recall $d$-majorization. Take a distribution $d$ over a finite set $\Omega$ that is strictly positive everywhere, that is, $\supp(d)=\Omega$. (The contrary situation is non-physical, since it involves infinite energies by \eqref{gibbs distri}.)
In order to define a relation $\preceq_d$ between distributions that captures the notion of being close to a non-uniform distribution $d$, the basic idea is to use a map $\Lambda_d$ that takes $d$ to a uniform distribution in a larger space and maps the other distributions into that space respecting their distance relative to $d$. Then, the relation between these distributions is given by majorization on their image through $\Lambda_d$. We provide the details in the following paragraph.   

Assume $d_i \in \mathbb Q$ for all $i \in \Omega$ (otherwise, the desired result follows from a limit via rational distributions) and consider some $\alpha \in \mathbb R$ such that $\alpha d_i \in \mathbb N$ for all $i \in \Omega$. We refer to the larger space where the distributions will be mapped to by $\Omega_d'$, where $|\Omega_d'| \coloneqq \sum_{j=1}^{|\Omega|} |A_j|$ and $|A_j| \coloneqq \alpha d_j$ for $1 \leq j \leq |\Omega|$. The map we consider is $\Lambda_d: \mathbb P_{\Omega} \to \mathbb P_{\Omega_d'}$, where, for all $p \in \mathbb P_{\Omega}$,
\begin{equation*}
    (\Lambda_d p)_i \coloneqq \frac{1}{\alpha} \frac{p_j}{d_j} \text{ } \forall i \in A_j,
\end{equation*}
for all $j$ such that $1 \leq j \leq |\Omega|$ \cite{brandao2015second,lostaglio2019introductory,gottwald2019bounded}. We can now define $d$-majorization. We say $q$ $d$-majorizes or thermo-majorizes $p$ and write $p \preceq_d q$ if and only if $(\Lambda_d p) \preceq (\Lambda_d q)$.
We refer to the model of relative uncertainty based on $d$-majorization $\preceq_d$ as
 \emph{$d$-disorder} or \emph{disorder relative to $d$} and define by analogy with majorization a \emph{family of second laws of $d$-disorder}.

 Now that we have defined $d$-majorization, we address the same question as in Section \ref{main sect} but for $d$-majorization, that is, we ask whether families of second laws of $d$-disorder exist, and if so, what the minimal number of members of such a family is. Since the situation is analogous to that of majorization for the cases where $|\Omega| \geq 3$ and significantly different whenever $|\Omega|=2$, we consider first the former in the following section.

 \subsection{Second laws of d-majorization with \texorpdfstring{$|\Omega| \geq 3$}{TEXT}}

 We show in this section that $d$-disorder behaves like disorder regarding the second law provided $|\supp (d)| \geq 3$. In particular, in the following theorem, we show that a countably infinite family of second laws is required in this scenario. Recall that, given $d \in \mathbb P_\Omega$ such that $\supp (d) = \Omega$ and $p \in \mathbb P_\Omega$, the \emph{Kullback-Leibler divergence} \cite{cover1999elements} is defined as follows:
 \begin{equation*}
     D_\mathrm{KL}(p||d) \coloneqq \sum_{i=1}^{|\Omega|} p_i \log \Bigl(\frac{p_i}{d_i}\Bigr).
 \end{equation*}
 

\begin{theorem}
\label{d-majo dim} 
If $|\Omega| \geq 3$ and $d \in \mathbb P_\Omega$ with $\supp (d)=\Omega$, then the smallest family of second laws of $d$-disorder is countably infinite.
\end{theorem}

\begin{proof}
For simplicity, we assume $d$ to be non-increasingly ordered. Otherwise, a simple permutation of all the involved distributions yields the desired result. Moreover, we use $n \coloneqq |\Omega|$.

In order to show the result, we follow an analogous approach to the one in the proof of Theorem \ref{dim majo}. In particular, we show that countably infinite families of second laws exist and construct, for each $d$, a set of probability distributions $S_d \subseteq \mathbb P_\Omega$ such that $(S_d,\preceq_{S_d})$ is order isomorphic to $(X,\preceq')$ (which we defined via \eqref{simple partial} in Theorem \ref{dim majo}), where $\preceq_{S_d}$ is the restriction of $\preceq_d$ on $S_d$. We conclude, following Theorem \ref{dim majo}, that the smallest family of second laws of $d$-disorder is countably infinite whenever $|\Omega| \geq 3$

It is direct to show that a countably infinite family of second laws exist. We can simply follow the argument in Theorem \ref{dim majo} substituting \smash{$(u_i)_{i=1}^{|\Omega|-1}$} by the finite family 
\begin{equation*}
    (u_i \circ \Lambda_d)_{i=1}^{|\Omega_d'|-1}
\end{equation*}
and negative entropy by the Kullback-Leibler divergence.

To conclude, we construct $S_d$ and show it is order isomorphic to $(X,\preceq')$. Fix $\varepsilon, \gamma,m >0$ such that
$\varepsilon < d_n$, $m \coloneqq \varepsilon \frac{z}{1+z}$ with $z \coloneqq \frac{d_1}{d_2}$
and $\gamma < \varepsilon - m$, and take
\begin{equation*}
\begin{split}
S_d &\coloneqq \{p_x, q_x\}_{x \in (d_1+m,d_1 + \varepsilon - \gamma)}, \text{ where} \\
 p_x &\coloneqq (x, d_2+ \varepsilon-(x- d_1), d_3,\dots,d_{n-1},d_n-\varepsilon) \text{ and} \\
 q_x &\coloneqq (x, d_2+ \varepsilon-(x- d_1)-\gamma, d_3,\dots,d_{n-1},d_n-\varepsilon+\gamma)
\end{split}
\end{equation*}
for all $x \in (d_1+m,d_1 + \varepsilon - \gamma)$.
To obtain that $(S_d,\preceq_{S_d})$ is order isomorphic to $(X,\preceq')$, it suffices to show that, for any pair $x,y \in (d_1+m,d_1 + \varepsilon - \gamma)$ with $x <y$, we have $p_x \prec_d p_y$, $q_x \prec_d q_y$, $q_x \prec_d p_x$ and $p_x \bowtie_d q_y$. In order to check these relations, we first notice that, for all $r \in S_d$, we have \smash{$(\Lambda_d r)^{\downarrow}_j = r_{i}/(\alpha d_{i})$} for all $j$ such that $\alpha \sum_{k=1}^{i-1} d_k< j \leq \alpha \sum_{k=1}^{i} d_k$. To see this, it suffices to consider some $p_x$ ($q_x$ follows analogously) with $x \in (d_1+m,d_1 + \varepsilon - \gamma)$. In particular, we only need to establish that $x/d_1 > (d_2+ \varepsilon -(x-d_1))/d_2>1$, given that the other components of $\alpha (\Lambda_d p_x)$ are lower or equal to one. The first inequality is equivalent to $x>d_1+\frac{z}{1+z} \varepsilon$, which is true by definition of $S_d$, while the second is true since $x < d_1 + \varepsilon$ also by definition (in the case of $q_x$, it follows since, actually, $x < d_1 + \varepsilon - \gamma$). Now that we know how the components in the distributions belonging to $S_d$ are ordered after applying $\Lambda_d$, we can check the desired relations. $p_x \prec_d p_y$ (and, equivalently, $q_x \prec_d q_y$) follows from the fact that
$u_i(\Lambda_d p_x)<u_i(\Lambda_d p_y)$ whenever $1 \leq i <\alpha (d_1+d_2)$ and $u_i(\Lambda_d p_x)=u_i(\Lambda_d p_y)$ if $\alpha (d_1+d_2) \leq i < \alpha$.
$q_x \prec_d p_x$ follows from the fact that
$u_i(\Lambda_d q_x)=u_i(\Lambda_d p_x)$ if $1 \leq i \leq \alpha d_1$ and $u_i(\Lambda_d q_x)<u_i(\Lambda_d p_x)$ whenever $\alpha d_1 < i < \alpha$.
Lastly, $p_x \bowtie_d q_y$ since $u_1(\Lambda_d p_x)=x/(\alpha d_1) < y/(\alpha d_1)=u_1(\Lambda_d q_y)$ and $u_{\alpha (d_1+d_2)}(\Lambda_d q_y)=d_1+d_2+ \varepsilon - \gamma < d_1+d_2+ \varepsilon=u_{\alpha (d_1+d_2)}(\Lambda_d p_x)$.
\end{proof}

Contrasting Theorems \ref{dim majo} and \ref{d-majo dim} we see that, if $|\Omega| \geq 3$, then the inclusion of a non-constant energy function does not alter the minimal number of functions constituting a family of second laws. As we will see, this similarity does not hold for $|\Omega| = 2$. Now that we have dealt with the cases where $|\Omega| \geq 3$, we consider $|\Omega| = 2$ in the following section.

\subsection{Second laws of d-majorization with \texorpdfstring{$|\Omega| = 2$}{TEXT}}

Despite their similarity whenever $|\Omega| \geq 3$, majorization and $d$-majorization present substantial differences regarding the number of functions needed in a family of second laws when $|\Omega|=2$. In this scenario, as we stated in \eqref{second law dim2}, Shannon entropy constitutes a family of second laws of disorder. It is easy to see that a family only consisting of one function does not always exist for $d$-disorder when $|\Omega|=2$. In order to do so, it suffices to notice that 
$\preceq_d$ is not always \emph{total}, that is, that there exist distributions $p,q \in \mathbb P_\Omega$ that are incomparable.
Taking, for instance, $d_0=(2/3,1/3)$, it is easy to see that $p \bowtie_{d_0} q$, where $p=(11/12,1/12)$ and $q=(1/3,2/3)$. Despite one function not being always enough, one may think that a finite family of them may be. However, as we show in the following theorem, this is never the case.


\begin{theorem}
\label{thm: small state space}
    If $|\Omega| = 2$, $d \in \mathbb P_\Omega$ is not the uniform distribution and 
    $\supp (d)=\Omega$,
    then the smallest family of second laws of $d$-disorder is countably infinite.
\end{theorem}

\begin{proof}
    We fix here $d=(d_0,1-d_0)$. Since $d$ is not the uniform distribution, $|\supp (d)|=2$ and we can assume w.l.o.g. it is not-increasingly ordered, we have $1-d_0<d_0<1$ and, hence, $1/2<d_0<1$.
    
    In order to show the result, we begin noticing that the existence of a countable family of second laws of $d$-disorder follows exactly as in Theorem \ref{d-majo dim}. To conclude the proof, we ought to see that families of second laws of $d$-disorder with less functions do not exist. In order to do so, it suffices to characterize $(\mathbb P_\Omega, \preceq_d)$. Once this is done, we get that $(\mathbb P_\Omega, \preceq_d)$ has an order-theoretic structure that is quite close to that of $(X,\preceq')$ in Theorem \ref{dim majo}. We can then argue along the lines in \cite[Theorem 3 $(ii)$]{hack2022geometrical} that the smallest family of second laws of $(\mathbb P_\Omega, \preceq_d)$ is countably infinite.

    We begin, thus, characterizing $(\mathbb P_\Omega, \preceq_d)$. By definition, given some $p=(p_0,1-p_0) \in \mathbb P_\Omega$, the key parameter to establish its relation to other distributions in terms of $\preceq_d$ is the relation between $p_0$ and $d_0$ in terms of $\leq$, since that determines the $\leq$-order of the components of $\Lambda_d p$. In particular, if $p_0=d_0$, then $p=d$ and $p \preceq_d q$ for all $q \in \mathbb P_\Omega$. If $p_0>d_0$, then $p_0/d_0>(1-p_0)/(1-d_0)$ and, if  $p_0<d_0$, then $p_0/d_0<(1-p_0)/(1-d_0)$. To conclude the characterization of $(\mathbb P_\Omega, \preceq_d)$, we find the explicit $\preceq_d$-relation for any pair $p,q \in \mathbb P_\Omega$, taking $q=(q_0,1-q_0)$ and distinguishing different cases depending on the relation of their first components with $d_0$.
We consider the following cases:
\begin{enumerate}[label=(\roman*)]
\item  In case $d_0<q_0 \leq p_0$, then it is easy to see that $q \preceq_d p$.
\item Moreover, if $p_0 \leq q_0 <d_0$, then we also have that $q \preceq_d p$.
\item Consider now the case where $q_0<d_0<p_0$. In this scenario, if $(1-q_0)/(1-d_0) \leq p_0/d_0$, then it is easy to see that $q \preceq_d p$. The situation is richer in case $(1-q_0)/(1-d_0) > p_0/d_0$. In particular, one can show that $p \preceq_d q$ holds except in the case where $(1-q_0) + q_0(2d_0-1)/d_0 <p_0$. (This is the case since, if one of the non-increasing partial sums gives a larger value to $p$, then the non-increasing partial sum that attains the value $p_0$ for $p$ must also give $p$ a larger value.) Furthermore, the condition $(1-q_0) + q_0(2d_0-1)/d_0 <p_0$ holds if and only $1 + q_0(d_0-1)/d_0 < p_0$. In this scenario, we have that $p \bowtie_d q$. Moreover, for each $q =(q_0,1-q_0) \in \mathbb P_\Omega$ such that $q_0<d_0$, there exists a continuum of $p =(p_0,1-p_0) \in \mathbb P_\Omega$ fulfilling these conditions. This is the case since we have $d_0<1 + q_0(d_0-1)/d_0<(1-q_0)d_0/(1-d_0)$. (One can see this by manipulating both inequalities that they are both equivalent to $q_0<d_0$, which is true for the case we are considering.)  This concludes the characterization of $(\mathbb P_\Omega, \preceq_d)$.
\end{enumerate}

    \begin{figure}[!tb]
\centering
\begin{tikzpicture}
    \node[main node] (1) {$p_x$};
    \node[main node] (2) [right = 2cm  of 1]  {$p_y$};
    \node[main node] (3) [right = 2cm  of 2]  {$p_z$};
    \node[main node] (4) [below = 2cm  of 1] {$q_x$};
    \node[main node] (5) [right = 2cm  of 4] {$q_y$};
    \node[main node] (6) [right = 2cm  of 5] {$q_z$};
    \node[main node] (7) [left = 2cm  of 4] {$d$};

    \path[draw,thick,->]
    (7) edge node {} (1)
    (7) edge node {} (4)
    (1) edge node {} (6)
    (4) edge node {} (5)
    (5) edge node {} (6)
    (1) edge node {} (2)
    (2) edge node {} (3)
    (4) edge node {} (1)
    (6) edge node {} (3)
    (5) edge node {} (2)
    ;
\end{tikzpicture}
\caption{Representation of $(\mathbb P_\Omega, \preceq_d)$ when both $|\Omega|=2$ and $d$ is not a uniform distribution. We include $d$ and six elements in $S_{d_0}$, which we defined in \eqref{last set def} and used in Theorem \ref{thm: small state space} to show that no finite family of second laws of $d$-disorder exists provided $|\Omega|=2$ and $d$ is not a uniform distribution.
The elements in $S_{d_0}$ that we include are $q_r=(r,1-r)$ and $p_r=((1-r)d_0/(1-d_0),1-(1-r)d_0/(1-d_0))$ for $r \in \{x,y,z\}$. Moreover, in order to assure that the order relations are precisely the ones we indicate (plus the ones given by transitivity), we fix some $z$ such that $t<z<d_0$, take $x=(2d_0-1)/d_0+z(1-d_0)^2/d_0^2$ (it is easy to see that $x < d_0$) and take some $y$ such that $z<y<x$. Note that, if $d$ were uniform, then we would have $q_x \sim_d p_x$ for all $x \in (t,d_0)$ and, hence, $S_{d_0}$ would have no incomparable distributions (cf. Figure \ref{uniform dim 2})}.
\label{fig:small state space}
\end{figure}
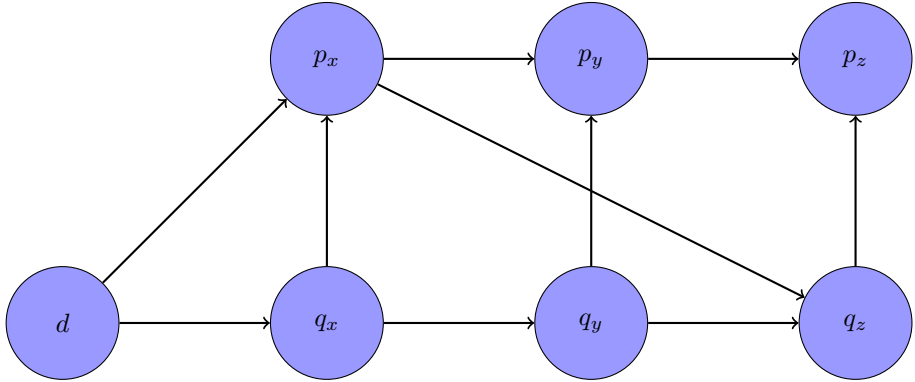

Now that we have characterized $(\mathbb P_\Omega, \preceq_d)$, it suffices to find a subset of $(\mathbb P_\Omega, \preceq_d)$ that resembles $(X,\preceq')$ from Theorem \ref{dim majo} to conclude the proof. In particular, we take
some $t \in \mathbb R$ such that $(2d_0-1)/d_0 \leq t <d_0$ (we can always do this since $d_0^2-2d_0+1=(d_0-1)^2 >0$), and we consider the set 
    \begin{equation}
    \label{last set def}
    \begin{split}
S_{d_0} &\coloneqq \{p_x, q_x\}_{x \in (t,d_0)}, \text{ where} \\
 p_x &\coloneqq ((1-x)d_0/(1-d_0), 1-(1-x)d_0/(1-d_0)) \text{ and} \\
 q_x &\coloneqq (x, 1-x)
\end{split}
\end{equation}
for all $x \in (t,d_0)$. Hence, we have, for each $q_x \in S_{d_0}$, that there exists some $y \in (t,d_0)$ such that $q_x \preceq_d p_y$ and $q_x \bowtie_d p_z$ for an open interval inside $(t,d_0)$ for which $y$ is an endpoint. Thus, if we denote by $\preceq_{S_{d_0}}$ the restriction of $\preceq_d$ on $S_{d_0}$, 
we have that $(S_{d_0},\preceq_{S_{d_0}})$
behaves quite similarly to $(X,\preceq')$ and we can argue essentially in the same way as in \cite[Theorem 3 $(ii)$]{hack2022geometrical} that any family of second laws for $(S_{d_0},\preceq_{S_{d_0}})$ is, at least, countably infinite. This concludes the proof. (We include a representation of $(\mathbb P_\Omega, \preceq_d)$ in Figure \ref{fig:small state space}, where we also include some elements of $S_{d_0}$.)
\end{proof}

Theorem \ref{thm: small state space} shows that the inclusion of a non-constant energy function can have dramatic consequences regarding the number of functions required to form a family of second laws provided $|\Omega|=2$. In particular, whenever there is no distinctions between compartments, then the order relation is total and a single function can constitute a family of second laws \eqref{second law dim2}. However, whenever the energy functions is non-constant, transitions are driven by both energy and randomness. Intuitively, we can think of the lack of agreement between these two objective functions as causing pairs of distributions to be unrelated by $d$-majorization. This directly implies that no function alone can constitute a family of second laws and, ultimately, leads to the non-existence of finite families of second laws. This contrasts with the classical situation in thermodynamics, where energy and entropy are integrated into a single function, the \emph{free energy}, whose value determines whether transitions are possible.

It is well-known \cite{lostaglio2019introductory,ruch1978mixing,ruch1980generalization} that, for any pair $p,q \in \mathbb P_\Omega$,  $q \preceq_d p$ if and only if there exists a \emph{Gibbs stochastic} matrix $M$ (i.e. a stochastic matrix such that $Md=d$ for some Gibbs distribution $d \in \mathbb P_\Omega$) such that $q = M p$. Hence, the basic distinction between the constant and non-constant energy scenarios for $|\Omega|=2$ can be traced back to the fact that, if $d \in \mathbb P_\Omega$ is uniform and $p,q \in \mathbb P_\Omega$, then there is always a Gibbs stochastic matrix (which is usually called \emph{doubly stochastic} whenever $d$ is uniform) $M$ such that either $p=Mq$ or $q=Mp$, which is no longer the case for any non-uniform $d$.

It is also worth mentioning that, for $|\Omega|=2$, the cardinality of the minimal families of second laws are highly 
non-continuous in $d$. That is, if $u \in \mathbb P_\Omega$ is the uniform distribution over $\Omega$ and $(d_n)_{n \geq 0} \subseteq \mathbb P_\Omega$ is a sequence of distributions converging to $u$, $\lim_{n \to \infty} d_n = u$, such that $|\supp (d_n)|=2$ and $d_n \neq u$ for all $n \geq 0$, then 
\begin{equation}
\label{discontinuous second law}
  \min_{(f_i)_{i \in I} \in SL(d_n)} |I| = |\mathbb N| > 1 = \min_{(f_i)_{i \in I} \in SL(u)} |I|
 \end{equation}
 for all $n \geq 0$, where $SL(p)$ denotes the set of families of second laws for $p \in \mathbb P_\Omega$.

 Using the characterization of $(\mathbb P_\Omega,\preceq_d)$ in the proof of Theorem \ref{thm: small state space}, we can show that, if $|\Omega|=2$, then the lack of continuity for families of second laws in \eqref{discontinuous second law} also appears (although in a weaker form) when we consider multi-utilities (see \eqref{multiutil}). In particular, if $d$ is uniform, then multi-utilities consisting of a single function exist by \eqref{second law dim2}. However, in the same setting as \eqref{discontinuous second law}, we have
 \begin{equation}
\label{discontinuous multi-ut}
  \min_{(f_i)_{i \in I} \in MU(d_n)} |I| = 2 > 1 = \min_{(f_i)_{i \in I} \in MU(u)} |I|
 \end{equation}
 for all $n \geq 0$, where $MU(p)$ denotes the set of multi-utilities for $p \in \mathbb P_\Omega$. We obtain that \eqref{discontinuous multi-ut} holds as a direct consequence of the following proposition.
 \begin{proposition}
 If $|\Omega| = 2$, $d \in \mathbb P_\Omega$ is not the uniform distribution and
    $\supp (d)=\Omega$,
    then the smallest multi-utility for $(\mathbb P_\Omega,\preceq_d)$ consists of two functions. 
\end{proposition}
\begin{proof}
Fix w.l.o.g. $d=(d_0,1-d_0)$ with $d_0>1-d_0$. By the characterization in the proof Theorem \ref{thm: small state space}, we know that $(\mathbb P_\Omega, \preceq_d)$ has incomparable pairs of distributions. Hence, there are no multi-utilities consisting of a single function. Thus, in order to conclude the proof, it suffices to find a multi-utility consisting of two functions. In particular, we will show that $\mathcal U \coloneqq \{u,v\}$ is a multi-utility, where
\begin{equation}
\label{mult-ut 1}
    u(r)\coloneqq \begin{cases}
    1-r_0 \frac{1-d_0}{d_0} &\text {if } d_0 > r_0 ,\\
    r_0 &\text {if } d_0 \leq r_0,
    \end{cases}
\end{equation}
and 
\begin{equation}
\label{mult-ut 2}
    v(r)\coloneqq \begin{cases}
   \left(1-r_0\right) \frac{d_0}{1-d_0} &\text {if } d_0 > r_0 ,\\
    r_0 &\text {if } d_0 \leq r_0.
    \end{cases}
\end{equation}
for all $r=(r_0,1-r_0) \in \mathbb P_\Omega$.
We conclude the proof by showing that $\mathcal U$ is indeed a multi-utility. By the characterization in Theorem \ref{thm: small state space}, it is immediate to see that $u$ and $v$ are monotonic. Consider now a pair $p=(p_0,1-p_0),q = (q_0.1-q_0) \in \mathbb P_\Omega$ such that $\neg(p \preceq q)$. (Note that this implies $p,q \neq d$.) We begin taking care of the cases where $q \prec p$. If $q \prec p$ and either $p_0 \leq q_0 < d_0$ or $d_0 < q_0 \leq p_0$ holds, then $u(q)<u(p)$ and $v(q)<v(p)$. If $q \prec p$ and $q_0 < d_0 < p_0$ hold, then $v(q)<v(p)$. Lastly, if $q \prec p$ and $p_0 < d_0 < q_0$ hold, then $u(q)<u(p)$. To conclude, assume that $q \bowtie p$. In this scenario, we can fix w.l.o.g. $q_0<d_0<p_0$ and $1-q_0 (1-d_0)/d_0 < p_0 < (1-q_0)d_0/(1-d_0)$. Hence, $u(q)< u(p)$ and $v(q)>v(p)$. This concludes the proof. 
\end{proof}

 Note that, if $d$ is uniform, then the functions in \eqref{mult-ut 1} and \eqref{mult-ut 2} become equal to each other. Moreover, they become equal to $u_1$, which constitutes a single-function multi-utility according to \eqref{majo}.

In the following section, we return to our molecular diffusion example to apply our main results and to comment on the limitations regarding majorization in this instance. Moreover, we show how our results can be used when dealing with a conjecture regarding catalytic majorization.

 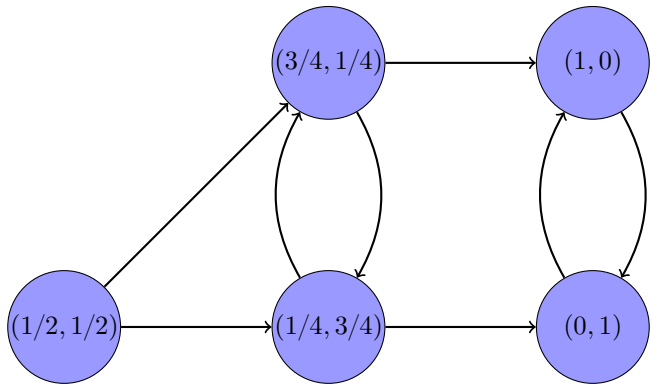
\begin{figure}[!tb]
\centering
\begin{tikzpicture}
    \node[main node] (1) {$(3/4,1/4)$};
    \node[main node] (2) [right = 2cm  of 1]  {$(1,0)$};
    \node[main node] (4) [below = 2cm  of 1] {$(1/4,3/4)$};
    \node[main node] (5) [right = 2cm  of 4] {$(0,1)$};
    \node[main node] (7) [left = 2cm  of 4] {$(1/2,1/2)$};

    \path[draw,thick,->]
    (7) edge node {} (1)
    (7) edge node {} (4)
    (4) edge node {} (5)
    (1) edge node {} (2)
    (1) edge [bend left] node {} (4)
    (4) edge [bend left] node {} (1)
    (2) edge [bend left] node {} (5)
    (5) edge [bend left] node {} (2)
    ;
\end{tikzpicture}
\caption{Representation of the majorization ordering for $|\Omega|=2$.}
\label{uniform dim 2}
\end{figure}

\section{Applications}
\label{application}

\subsection{Molecular diffusion}
\label{molec diff}

We return in this section to molecular diffusion. After commenting on
the limitations of modelling molecular diffusion via majorization, we introduce a variation of majorization that better captures this phenomenon and show the applicability of our results in the new model.

First of all, it should be noted that majorization clearly does not completely capture molecular diffusion. For example, if $|\Omega|=3$, majorization allows a transition from $q=(1/2,1/2,0)$
to $p=(1/2,0,1/2)$, which would clearly not take place by molecular diffusion. In fact, although majorization contains all the transitions which are permitted by molecular diffusion, it also contains some which are not. As we know from the study of elementary thermal operations (ETO) \cite{arnold2018majorization,lostaglio2018elementary,hack2023universality}, this is the case since
majorization can be characterized by permutations and the interchange of probability mass between (not necessarily adjacent) compartments,
which are both not part of molecular diffusion in general.
Moreover, majorization allows the permutation of probabilities between compartments, which is in general not allowed by molecular diffusion.

In order to avoid the difficulties stated in the previous paragraph, we can define a new order relation, in the spirit of majorization.



\begin{defi}[Molecular diffusion ordering]
\label{molec def}
Given some finite number of compartments $\Omega$, we say a distribution $p \in \mathbb P_\Omega$ can be \emph{achieved by molecular diffusion} from another distribution $q \in \mathbb P_\Omega$, and write $p \preceq_{md} q$, whenever there exists some $N \geq 1$ and a couple of finite families $(M_k)_{k=1,\dots,N}$, $(\lambda_{j,k})_{k=1,\dots,N; j=0,\dots,M_k-1}$ with $M_k \geq 1$ and $0 \leq \lambda_{j,k} \leq 1$ for all $j,k$ such that 
\begin{equation*}
    p= \prod_{k=1}^N G_k q,
\end{equation*}
where 
\begin{equation}
\label{nearest neigh}
    G_k= \lambda_{0,k} \mathbb I + \sum_{j=1}^{M_k-1} \lambda_{j,k} P(i_{l_{j,k}},i_{l_{j,k}}+1)
\end{equation}
and $\sum_{j=0}^{M_k-1} \lambda_{j,k} = 1$ for all $k$, $1 \leq l_{j,k}<|\Omega|$ for all $j,k$, $\mathbb I$ is the identity matrix, and $P(s,t)$ denotes the matrix
that, for all $r \in \mathbb P_\Omega$, equates entries $s$ and $t$
\begin{equation}
\label{diffusion element}
    (r_s,r_t) \mapsto \left( (P(s,t)r)_s, (P(s,t)r)_t \right) = \left(\frac{1}{2} (r_s+r_t), \frac{1}{2} (r_s+r_t) \right)
\end{equation}
and acts as the identity on the other components.
\end{defi}

Note that the molecular diffusion ordering corresponds to the so-called \emph{diffusion model} on the graph $P_{|\Omega|}$, which has been studied by the plasma physics community \cite{hay2017extreme,kolmes2020recovering}. The solution to the issues presented before its definition comes from the fact only interactions between adjacent compartment are allowed by \eqref{nearest neigh} and that the permutation of probabilities is prohibited by \eqref{diffusion element}. Definition \eqref{molec def} resembles elementary thermal operations, except for the fact that ETO usually allow the interaction between any pair of compartments and, moreover, the permutation of probabilities is allowed (cf. \cite{lostaglio2018elementary}). Definition \eqref{molec def} is also closely related to \emph{continuous majorization} \cite{lostaglio2022continuous,korzekwa2022optimizing,zylka1985note}, which avoids the permutation of probabilities by adding a continuity constraint to majorization (see \cite[Theorem 3]{lostaglio2022continuous}) but allows the interaction between non-adjacent compartments.

Note that the uniform distribution remains unchanged when any operator of the form prescribed by Definition \eqref{molec def} is applied (in particular, $(r_s,r_t) \mapsto (r_s,r_t)$ provided $r_s=r_t=1/|\Omega|$ in \eqref{diffusion element}), and that $p \preceq_{md} q$ implies $p \preceq q$ since any operator allowed by \eqref{molec def} is a doubly stochastic matrix. Moreover, as expected, there is no steady state aside from the uniform distribution. (This is the case since any distribution with a couple of different entries $r_s \neq r_t$ will not be mapped to itself by \eqref{diffusion element}.)



By analogy with majorization, we can define families of second laws of molecular diffusion. In fact, the main result in this work still holds, namely, an infinite family of functions is needed in order to emulate the role of the second law for molecular diffusion provided $|\Omega| \geq 3$.
The reason for this is that the transitions which are not allowed by majorization do not happen by molecular diffusion and, in particular, the transitions allowed by majorization in $S \subseteq \mathbb P_{\Omega}$ (see the proof of Theorem \ref{dim majo}) would also take place by molecular diffusion given that they only involve the exchange of probability mass between adjacent elements in $\Omega$ (which correspond to the interchange of molecules between adjacent compartments in our example with the gas in a box). Hence, the order relations in $S$ given by majorization coincide with those given by molecular diffusion and the conclusion also holds in this case. The case where $|\Omega|=2$ differs with the single-function scenario  from majorization and actually requires at least two functions to build a family of second laws. We show this in the following corollary.



\begin{corollary}
    If $|\Omega| \geq 3$, then there is no finite family of second laws of molecular diffusion. However, if $|\Omega| = 2$, then the smallest family of second laws of molecular diffusion consists of two functions.
\end{corollary}

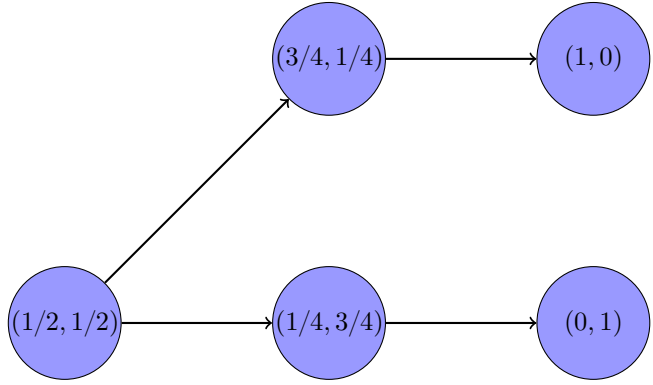
\begin{figure}[!tb]
\centering
\begin{tikzpicture}
    \node[main node] (1) {$(3/4,1/4)$};
    \node[main node] (2) [right = 2cm  of 1]  {$(1,0)$};
    \node[main node] (4) [below = 2cm  of 1] {$(1/4,3/4)$};
    \node[main node] (5) [right = 2cm  of 4] {$(0,1)$};
    \node[main node] (7) [left = 2cm  of 4] {$(1/2,1/2)$};

    \path[draw,thick,->]
    (7) edge node {} (1)
    (7) edge node {} (4)
    (4) edge node {} (5)
    (1) edge node {} (2)
    ;
\end{tikzpicture}
\caption{Representation of the molecular diffusion ordering for $|\Omega|=2$. In this scenario, the only order relations are, for all $p \in \mathbb P_\Omega$, $(\frac{1}{2},\frac{1}{2}) \preceq_{md} p$ and, for all $q \in \mathbb P_\Omega$, $p \preceq_{md} q$ whenever $p \preceq q$ and either $p_1 > p_2$ and $q_1 > q_2$ or $p_1 < p_2$ and $q_1 < q_2$ hold.}
\label{md dim 2}
\end{figure}

\begin{proof}
    If  $|\Omega| \geq 3$, then the result follows as a direct consequence of Theorem \ref{dim majo}. In particular, we simply have to show that $\preceq$ and $\preceq_{md}$ coincide on the set $S \subseteq \mathbb P_\Omega$ defined in \eqref{def S}. In order to do so, following the notation in the proof of Theorem \ref{dim majo}, it is enough to notice that, for any pair $x,y \in (\frac{1}{2}, \frac{1}{2} + \varepsilon - \gamma)$ with $x <y$, we have $p_x \prec_{md} p_y$, $q_x \prec_{md} q_y$, $q_x \prec_{md} p_x$ and $p_x \bowtie_{md} q_y$. $p_x \bowtie_{md} q_y$ is clear since, if that was not the case, then we would contradict the fact that $p_x \bowtie q_y$. $p_x \prec_{md} p_y$ since, by definition, we have $(p_y)_2<(p_x)_2<(p_x)_1<(p_y)_1$ and, hence, there exists some $\lambda$ such that $0<\lambda<1$ and $p_x= [(1-\lambda) \mathbb I + \lambda P(1,2)] p_y$. We obtain analogously that $q_x \prec_{md} q_y$. Lastly, since we have $(p_x)_3<(q_x)_3<(q_x)_2<(p_x)_2$, there exists some $\lambda$ such that $0<\lambda<1$ and $q_x= [(1-\lambda) \mathbb I + \lambda P(2,3)] p_x$. (Note that there are no permutation of probabilities in these cases.)
    

    If $|\Omega|=2$, then it is easy to see that the molecular diffusion ordering is given by Figure \ref{md dim 2}. Since there are incomparable pairs of distributions, like $p=(0,1)$ and $q=(1,0)$, any family of second laws of molecular diffusion must have, at least, two functions. Lastly, it is not hard to see that $\{h_1,h_2\}$ is a family of second laws of molecular diffusion, where
    \begin{equation*}
        h_1(p) \coloneqq  \begin{cases}
    p_1 &\text{if } p_2 \geq p_1,\\
    -p_1 &\text{otherwise,}
    \end{cases}
    \end{equation*}
    and
     \begin{equation*}
        h_2(p) \coloneqq  \begin{cases}
    p_2 &\text{if } p_1 \geq p_2,\\
    -p_2 &\text{otherwise,}
    \end{cases}
    \end{equation*}
     for all $p \in \mathbb P_\Omega$.
\end{proof}

\subsection{Trumping}
\label{sec trumping}

We can illustrate the applicability of our results here via a variation of majorization, known as \emph{trumping} or \emph{catalytic majorization} and denoted by $\preceq_T$, which has raised considerable interest recently \cite{jonathan1999entanglement,daftuar2001mathematical,turgut2007catalytic,klimesh2007inequalities,muller2016generalization}. Given a finite set $\Omega$ and a pair of distributions $p,q \in \mathbb{P}_{\Omega}$, trumping is defined as follows:
     \begin{equation}
     \label{def trump}
        p \preceq_T q \iff \exists r \in \mathbb{P}_{\Omega'} \text{ } (|\Omega'|< \infty) \text{ s.t. } p \otimes r \preceq q \otimes r,
\end{equation}
where $p \otimes r:= (p_1r_1,..,p_1r_{\Omega'},..,p_{\Omega}r_1,..,p_{\Omega} r_{\Omega'})$ for all $p \in \mathbb{P}_{\Omega}$, $r \in \mathbb{P}_{\Omega'}$. Note that trumping extends majorization though a third state $r$, known as the \emph{catalyst}.


The most extended characterization of trumping in terms of functions can be found in \cite{turgut2007catalytic} (see also \cite{klimesh2007inequalities}). There, aside from some details (see \cite[Theorem 1]{turgut2007catalytic}), it is shown that 
\begin{equation}
\label{trumping rel}
    p \prec_T q \iff f_r(p) < f_r(q) \text{ } \forall r \in \mathbb R,
\end{equation}
where $(f_r)_{r \in \mathbb R}$ is a family of (information-theoretic) real-valued functions, $f_r: \mathbb P_\Omega \to \mathbb R$. Note that the literature on trumping is written in terms of strict monotones instead of second laws. However, despite being part of resource theory, \eqref{trumping rel} characterizes irreversibilities in the same fashion as our family of second laws, going further in this regard than the more common multi-utilities. In fact, it is not hard to see that, for any preorder $\preceq$, a family of functions is a multi-utility and fulfills the equivalent of \eqref{trumping rel} if and only if it is a family of second laws for $\preceq$. We can, hence, think of our families of second laws as condensing the properties in the two real-valued representations that have been, to our best knowledge, considered in resource theory. 

Before we continue, it is worth mentioning a variation of trumping, \emph{correlated trumping}, whose relation with the second law has been considered in the literature \cite{muller2016generalization,muller2018correlating}. Instead of leaving the catalyst unaltered at the end of the process, correlated trumping allows the formation of correlations among $k \in \mathbb N$ auxiliary systems. Close to our notion of family of second laws are the characterizations of correlated trumping (with $k \geq 3$) via Hartley and Shannon entropies \cite[Theorem 1]{muller2016generalization} and the characterization of Helmholtz free energy in \cite[Main result 1]{muller2018correlating}.

Aside from the case with $|\Omega|=2$ (which is equivalent to majorization), it was conjectured in \cite{klimesh2004entropy} that no proper subset of $(f_r)_{r \in \mathbb R}$ suffices for
a characterization like the one in \eqref{trumping rel} to arise.
In the following corollary, we use Theorem \ref{dim majo} to show that no finite proper subset of $(f_r)_{r \in \mathbb R}$ suffices. 
In fact, we prove that, in general, no finite family of real-valued functions 
suffices to obtain a characterization like that in \eqref{trumping rel}.

\begin{corollary}
\label{trumping coro}
If $|\Omega| \geq 3$ and $(g_i)_{i \in I}$ is a family of functions, $g_i: \mathbb P_\Omega \to \mathbb R$ for all $i \in I$, such that, for all $p,q \in \mathbb P_\Omega$, we have 
\begin{equation}
\label{trumping rel 2}
    p \prec_T q \iff g_i(p) < g_i(q) \text{ } \forall i \in I,
\end{equation}
then $I$ cannot be finite.
\end{corollary}

\begin{proof}
    We will prove the result holds by reduction to the absurd. In particular, we will assume the statement holds for a finite family and, after showing the equivalence between majorization and trumping on the set $S$ defined in Theorem \ref{dim majo}, we will prove our assumption implies the existence of a finite family of second laws of disorder on $S$, which contradicts the statement in Theorem \ref{dim majo}.
    
    Assume, thus, there exists a family $(g_i)_{i \in I}$ with finite $I$ such that \eqref{trumping rel 2} holds. The first thing we ought to notice is that, for all $p,q \in S$ (where we take $S$ as defined in \eqref{def S}), majorization and trumping coincide, i.e.  $p \preceq q$ if and only if $p \preceq_T q$. Proving sufficiency is direct by definition of trumping. To prove necessity, we proceed by contrapositive. Take  a pair $p,q \in S$ such that $\neg(p \preceq q)$ and consider two cases. If $q \prec p$, then $q \prec_T p$ by definition of trumping  \cite[Theorem 3]{daftuar2001mathematical}, hence $\neg(p \preceq_T q)$. Moreover, if $q \bowtie p$, then, by definition of $S$, we either have $q_1<p_1$ and $q_3<p_3$ or $q_1>p_1$ and $q_3>p_3$. In either case we have $q \bowtie_T p$ by definition (as stated in \cite[Theorem 3]{daftuar2001mathematical}) and, hence, $\neg(p \preceq_T q)$. Lastly, since majorization and trumping coincide on $S$ as we just showed, we have that 
\begin{equation}
\label{trumping rel 3}
    p \prec q \iff g_i(p) < g_i(q) \text{ } \forall i \in I,
\end{equation}
for all $p,q \in S$.

To conclude the proof, it suffices to notice that \eqref{trumping rel 3} 
implies the existence of a finite family of second laws of disorder $(f_i)_{i \in I}$ that works, at least, on $S$, which yields the desired contradiction. To prove the last claim we need, we simply define define the family $(f_i)_{i \in I}$. We consider, in particular, $h_i(p) \coloneqq g_i(p^\downarrow)$ for all $p \in S$ and $i \in I$, and notice the following properties: If $p \sim q$, then $p^\downarrow = q^\downarrow$. Hence, $f_i(p)=f_i(q)$ for all $i \in I$. If $p \prec q$, then $p^\downarrow \prec q^\downarrow$. Hence, $f_i(p)<f_i(q)$ for all $i \in I$. If $p \bowtie q$, then there exist $y,z \in S$ such that $p \bowtie y \prec q$ and $q \bowtie z \prec p$ by definition of $S$. Thus, there exist $i,j \in I$ such that $f_i(p) \leq f_i(y)<f_i(q)$ and $f_i(q) \leq f_i(z)<f_i(p)$ hold. Hence, $(i)$, $(ii)$ and $(iii)$ hold for $(f_i)_{i \in I}$, where $f_i \coloneqq -h_i$ for all $i \in I$, and we reach a contradiction.
\end{proof}

\section{Conclusion}

In this paper, we have considered a fundamental notion of uncertainty, namely majorization, that possesses applications in several areas like thermodynamics or entanglement theory. We have discussed the sort of properties a family of functions ought to have in order to be considered as a second law for majorization and have shown that, whenever the state space is larger than two, any family of such functions is necessarily infinite. In particular, we can only use infinite families of functions to interpret majorization as the result of competing optimization principles. Hence, such a model goes well beyond the complexity of simply assuming the transitions to be exclusively determined by entropy, and we gain, thus, insight into the measurement complexity of Ruch's principle of increasing mixing character \cite{ruch1976principle}. Moreover, the study of majorization and its relation to real-valued functions may become a helpful tool in the ongoing study of generalized entropies and nonextensive statistical mechanics and thermodynamics (see, for example, \cite{tsallis2009introduction}).

As shown in Section \ref{molec diff}, our results are useful when considering the complexity of molecular diffusion. However, more generally, the description of transitions through order structures goes well beyond thermodynamics, where they were originally used. In fact, our results are helpful in the study of entanglement, where majorization models the transitions given by local operations and classical communications \cite{nielsen1999conditions}.

Our results have proven to be useful, as exemplified by the progress we made in Corollary \ref{trumping coro} on a conjecture concerning trumping \cite{turgut2007catalytic,klimesh2007inequalities}. They may continue assisting the study of trumping,
where issues regarding complexity are still to be settled. For example, only uncountable families of what we have called \emph{second laws} have been found there \cite{turgut2007catalytic}. Hence, given Corollary \ref{trumping coro}, it remains to be clarified whether countably infinite families of second laws exist.
The transitions in several other systems of interest, like general relativity \cite{bombelli1987space,surya2019causal}, have been defined in terms of order structures and, thus, would benefit greatly from a better comprehension in terms of measurements, like the one we have pursued here for majorization.

\newpage


\vskip2pc

\bibliographystyle{plain}

\bibliography{main} 

\end{document}